# Speaker Identification Experiments Under Gender De-Identification

Marcos Faundez-Zanuy, Enric Sesa-Nogueras
Grup de Tractament del Senyal
TecnoCampus Mataró-Maresme
Mataró, Spain
{faundez, sesa}@tecnocampus.cat

Stefano Marinozzi
Institute for Electronic Engeenering
Università Politecnica delle Marche
Ancona, Italy
stefano.marinozzi87@gmail.com

*Abstract*— The present work is based on the COST Action IC1206 for De-identification in multimedia content. It was performed to test four algorithms of voice modifications on a speech gender recognizer to find the degree of modification of pitch when the speech recognizer have the probability of success equal to the probability of failure. The purpose of this analysis is to assess the intensity of the speech tone modification, the quality, the reversibility and not-reversibility of the changes made.

*Keywords—De-Identification; Speech Algorithms*

## I. Introduction

In recent times many useful services have become available via web or over the telephone. With the increased usage of such services, the users are also becoming more aware of privacy implications of their use. Therefore applications that can assure that users can protect their privacy are becoming more attractive. Methods concerning person de-identification in still images or video which try to mask identification features such as faces, silhouettes, posture, gait etc. have already been proposed [1].

There is also a need for de-identification technologies in voice driven applications. For example, conversations may be recorded in call centres for various purposes, such analysis of operator mistakes, making the communication protocol more efficient in general or to prove a call has actually been made in case of complaints etc. In many cases, the identity of the caller is not important for the required purpose, and the customers may legitimately wonder why it should be recorded and kept. In [2] and [3] voice transformations were implemented to de-identify a small set of speakers and then tested with automatic speaker identification system. Voice transformation was successful in concealing identities of source speakers against the GMM-based speaker identification system. However in those experiments, speech samples from each person to be de-identified had to be available in advance in order to estimate the transformation parameters. In addition, those samples were parallel utterances, with the same text spoken by source and target speakers. Also to de-identify a speaker, his identity has to be known first, so that his corresponding voice transformation can be used for de-identification. This may also be a limitation in some cases, where the user doesn't want to identify with the system at all.

In a scenario with just a closed set of speakers to be de-identified, this may be acceptable, but in more realistic scenarios it would not be practical. In that case, the number of potential users of the system is extremely large, and many users will only use the system once. A requirement that the user has to supply would be inconvenient. For a practical application in such scenarios, it would be desirable that any new user can use the system immediately, without having to enrol with the system first in any way or identify himself, even for the purpose of de-identification.

In this paper, which extends one of our previous works on this topic [4], we propose two different approaches for voice modification: the first one is based on the Dolson and Laroche Phase Vocoder [5] the second one is based on the Vocal Tract Length Normalization (VTLN) which is based on the frequency warping factor [6], [7]. These approaches focus on modifying the non-linguistic characteristic of a given utterance without affecting its textual contents thus preserving the intelligibility of the spoken message.

State-of-the-art gender recognition systems nearly yield perfect recognition performances in clean conditions when using normal voices [8]. Using the threshold from this condition, we applied 25 steps of pitch modifications to all the BiosecurID database files [9] (15200 speech files), then, analyzed the response of the gender recognizer with respect to the pitch modification algorithm, for both masculine and feminine speech. Then, we created the graphics that will be discussed in a forthcoming section. Results show that the probability of success depends on the pitch modification algorithm and the gender of the speaker. To evaluate the performance (in voice quality) of the different algorithms, we carried out a listening test with 15 candidates, and we also compared machine performance with human performance.

## II. Voice modification algorithms

### A. The Phase Vocoder

The Phase Vocoder is presented as a high-quality solution for time-scale modification of signals, pitch-scale modifications usually being implemented as a combination of time scaling and sampling rate conversion. Pitch-shifting using the STFT representation of an audio signal as proposed in [10] is performed in four steps:

1. Peak detection: The simplest scheme consists of declaring that a bin is a peak if its magnitude is larger than that of its two (or four) nearest neighbours. It is assumed that each detected peak represents a sinusoidal component.

2. Define regions of influence: The region of influence is the sub-band around a spectral peak in which it is assumed that all phase values are dominated by the peak's phase. The boundaries of these sub-bands can be defined halfway between two peaks or at the lowest magnitude bin between two peaks.

3. Coefficient shift: Peaks and their regions of influence are shifted by frequency $\Delta\omega_m$, where $m$ is the peak index. If the relative amplitudes and phases of the bins around a sinusoidal peak are preserved during the translation, then the time-domain signal corresponding to the shifted peak is simply a sinusoid at a different frequency, modulated by the same analysis window.

4. Phase propagation: Since the frequencies of underlying sinusoids have been changed during the coefficient shift, phase-coherence from one frame to the next is lost. To avoid artefacts due to inter-frame phase inconsistency, phase values need to be updated.

The difference between the two types of Phase Vocoders considered lies in the function that describes the Phase propagation [11].

### B. The Vocal Tract Length Normalization (VTLN)

In order to change the voice into another, the spectrum of a frame has to be transformed [12]. Vocal tract length normalization is used to warp the spectrum of a frame, i.e., stretch or compress the spectrum with respect to the frequency axis which represents normalized frequencies in the range $0 \leq \omega \leq \pi$.

Frequencies are altered according to a warping function $g(\omega)$ which needs to be a monotonous function returning values between $0$ and $\pi$. Function $g(\omega)$ returns the warped position of the original frequency. Commonly, $g$ depends on a warping parameter affecting the shape of the function. The warping function is one of five predefined functions as shown in Table I. *Fig. 1* shows the corresponding illustration of these functions.

TABLE I. WARPING FUNCTION

| Type | Formula |
|---|---|
| Symmetric | $g(\omega, \alpha) = \begin{cases} \alpha\omega, & \omega \leq \omega_0 \\ \alpha\omega_0 + \dfrac{\pi - \alpha\omega_0}{\pi - \omega_0}(\omega - \omega_0), & \omega > \omega_0 \end{cases}$ $\omega_0 = \begin{cases} \dfrac{7\pi}{8}, & \alpha \leq 1 \\ \dfrac{7\pi}{8\alpha}, & \alpha > 1 \end{cases}$ |
| Asymmetric | $g(\omega, \alpha) = \begin{cases} \alpha\omega, & \omega \leq \omega_0 \\ \alpha\omega_0 + \dfrac{\pi - \alpha\omega_0}{\pi - \omega_0}(\omega - \omega_0), & \omega > \omega_0 \end{cases}$ $\omega_0 = \dfrac{7\pi}{8}$ |
| Quadratic | $g(\omega, \alpha) = \omega + \alpha\left(\left(\dfrac{\omega}{\pi}\right) - \left(\dfrac{\omega}{\pi}\right)^2\right)$ |
| Power | $g(\omega, \alpha) = \pi\left(\dfrac{\omega}{\pi}\right)^\alpha$ |
| Bilinear | $g(\omega, \alpha) = \left\|-i\dfrac{z - \alpha}{1 - \alpha z}\right\|$ $z = e^{i\omega}$ |

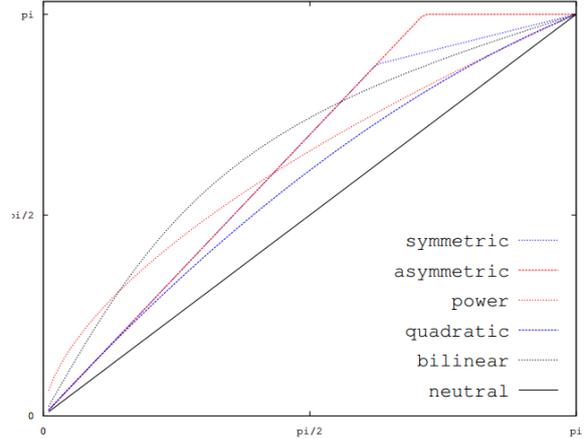

Fig. 1. Warping functions. The warping factor is 0.4 for the bilinear, 0.6 for the power function and 1.4 in all other cases

## III. Identification Algorithm

The identification algorithm is based on the arithmetic-harmonic sphericity measure [13]. A covariance matrix (CM) is computed for each speaker, and the following distance measure is applied:

$$\mu(C_j C_{test}) = \log\left(tr(C_{test} C_j^{-1}) tr(C_j C_{test}^{-1})\right) - 2\log(P) \quad (1)$$

where *tr* is the trace of the matrix.

The number of parameters for each speaker is ½($P^2$+P) (the covariance matrix is symmetric).

## IV. Experimental Results

We used speech samples from the BiosecurID database on an ASR trained whit normal voice and tested with disguised (modified) voice.
The voice modification applied using the Phase Vocoder is equivalent to the increase or decrease of the tone of the speech signal. The first one is called VOC and the second one is called VOCF. Regarding the VTLN, it depends on the two warping factors. For the Bilinear warp function incremental steps of 0.0065, for females, and decremental steps of 0.0043, for males, are applied. For the Quadratic warp function incremental steps of 0.057 for females and decremental steps of 0.029 for males are applied. *Fig. 2* and *Fig. 3* show the resulting trends for gender identification.

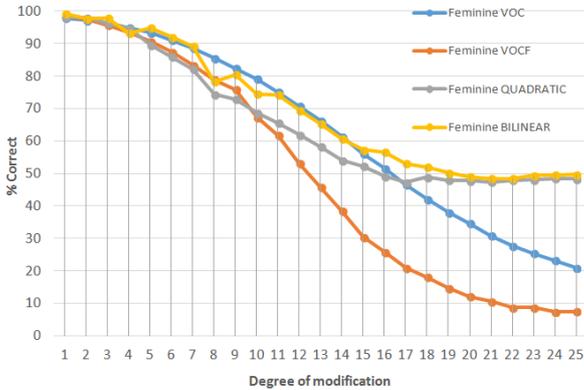

Fig. 2. Results for feminine voice modification

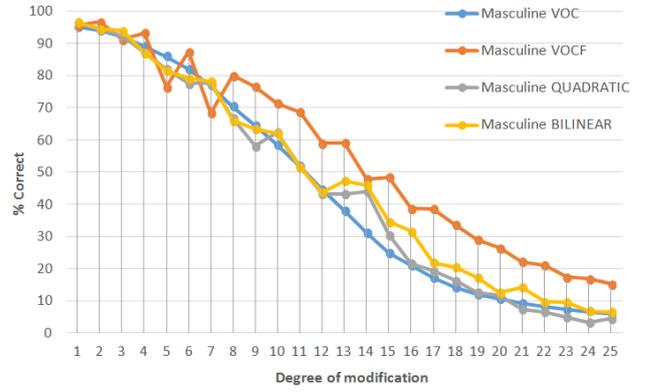

Fig. 3. Results for masculine voice modification

## V. Human Evaluation

De-identification with the purpose of not revealing the speaker's gender is only useful if the content of the transmitted information is still understandable for human listeners. The changes introduced by the algorithms can add noise and distortion which may cause some concern about the quality of the converted speech that becomes less natural and intelligible. Consequently, we conducted a human evaluation to investigate the intelligibility and to verify the gender of the speaker.

Our first test was on speaker identity. For every algorithm, we made 4 incremental changes (equivalent to the degree of modification 7, 13, 19, 25) for a total of 256 audio files.

Our listeners were asked to identify the speaker as a male or a female. The identification performed by human listeners is more accurate when compared to the gender recognition system: the results are similar for all the algorithms that we have used and while for the gender recognition system there is a general decay when the modification of the voice is stronger, the human identification requires a deeper change.

*Fig. 4*, *Fig. 5*, *Fig. 6* and *Fig. 7* show the trends for the four algorithms.

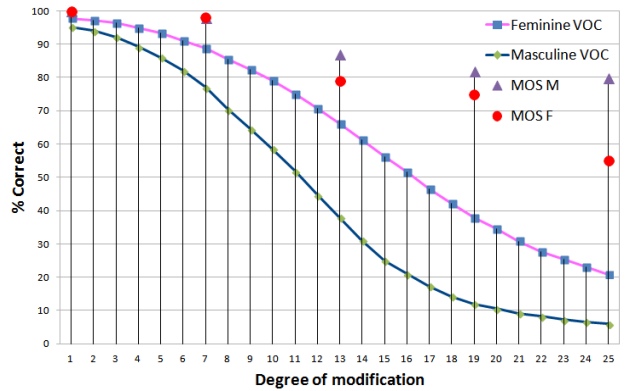

Fig. 4. MOS for VOC algorithm

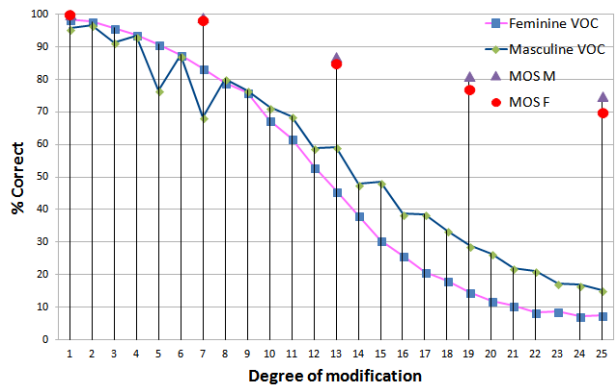

Fig. 5.  MOS for VOCF algorithm

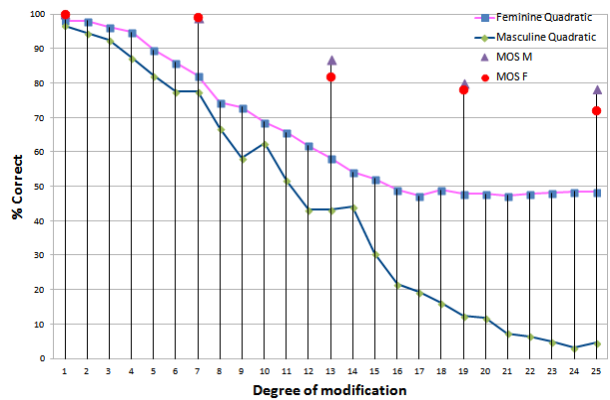

Fig. 6.  MOS for QUADRATIC algorithm

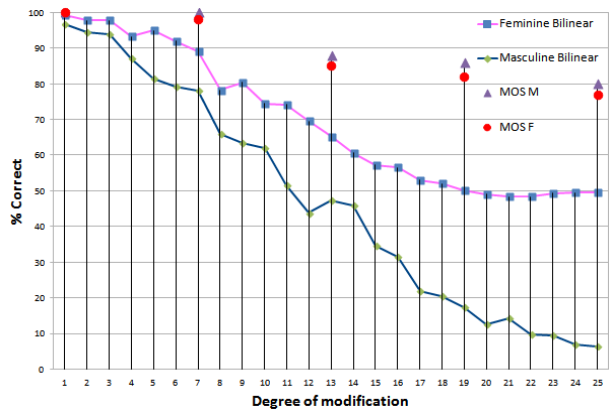

Fig. 7.  MOS for BILINEAR algorithm

The trends of the human evaluation are similar for all the algorithms.

The second test was on intelligibility. A successful de-identification process should preserve the intelligibility of the transmitted content. We played examples of the de-identified speech to listeners and asked them to assess the effort required to understand the meanings of sentences according to the following scale:

5 Complete relaxation possible; no effort required.
4 Attention necessary; no appreciable effort required.
3 Moderate effort required.
2 Considerable effort required.
1 No meaning understood with any feasible effort.

The results obtained are the following:
- Phase Vocoder (VOC): 3,7
- Phase Vocoder (VOCF): 3,3
- VTLN Quadratic: 3,4
- VTLN Bilinear: 3

Fig 8. shows the identification rate versus the degradation factor for the four algorithms.

Comparing figures 2 to 8 is evident that identity recognition drops faster than gender recognition when increasing the modification factor. This makes sense considering that gender recognition is a two-classes classification problem while identification is 1:N (N=400 in our experiments).

Another conclusion is that all the algorithms have a similar performance, with a tendency towards a null identification rate for modification rates equal or higher than 15.

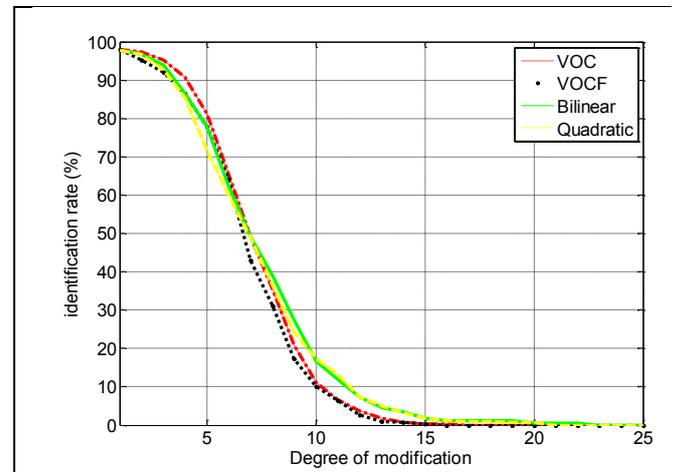

Fig. 8.  Identification rates

## VI. Conclusion and future work

In this paper, we studied the potential of voice transformation for gender de-identification. The method does not require enrolment of speakers for de-identification thus greatly extending possible applications of the system. We explored different voice transformation strategies including two kinds of Phase Vocoder that permit the reversibility of the changes made with a good intelligibility, and two functions of the Vocal Tract Length Normalization that not permit the reversibility of change made, keeping a sufficient intelligibility.

For the Phase Vocoder (VOC) we can see that when the error rate is at 50%, the degree of modification is 11 for male and 16 for female, while, for the Phase Vocoder (VOCF) the degree of modification is 13 for male and 16 for female.

For the VTLN with the Quadratic's warp function, the degree of modification is 11 for male and 16 for female, while, for the bilinear warp function, is 11 for male and 19 for female.

Further experiments were designed to test the ability of humans to recognize unknown speaker. In a carefully controlled experiment, human performance was measured and was compared to the gender recognition system. Results showed that machine performance is not comparable to average human performance.

In this paper we have checked that a good strategy for de-identification consists of a previous speech classification male/female in order to apply a different modification procedure. In addition our findings reveal that once gender de-identification is done, de-identification is not an issue.

We hope to expand this work in Automatic Speech Recognition research for children. The scope is to decrease the error rate using models trained on adult speech through the reduction of the pitch of the children speech.

# Acknowledgment

This work has been supported by FEDER and Ministerio de Ciencia e Innovación, TEC2012-38630-C04-03, COST IC-1206 and Erasmus Placement Mobility.